\documentclass[twocolumn,aps,showpacs]{revtex4}
\usepackage{graphicx}
\usepackage{graphics}
\begin{document}
\newcommand{\be}{\begin{equation}}
\newcommand{\ee}{\end{equation}}
\def\barr{\begin{array}}
\def\earr{\end{array}}
\newcommand{\ra}{\rightarrow}
\newcommand{\mr}{{\stackrel{<}{\sim}}}

\def\bi{\bibitem}
\def\lsim{\:\raisebox{-0.5ex}{$\stackrel{\textstyle<}{\sim}$}\:}
\def\gsim{\:\raisebox{-0.5ex}{$\stackrel{\textstyle>}{\sim}$}\:}
\def\gev{\; \rm  GeV}
\def\eg{ {\it e.g.}}
\def\ie{ {\it i.e.}}
\newcommand{\tb}{\tan \beta}

\title
{\begin{flushright}
IFT-2001/5 \\ [1.5ex]
{\large \bf hep-ph/0103223 }\\
updated version  \\ [1.5ex]
\end{flushright}
The new $(g-2)_{\mu}$ measurement and limits on the light Higgs bosons 
in 2HDM (II)}
\author{MARIA KRAWCZYK}
\affiliation{{\it Institute of Theoretical Physics, University of Warsaw,
ul. Ho\.za 69} 
{\it Warsaw, 00-681, Poland}}

\date{\today}

\begin{abstract}
We discuss how new  data for $a_{\mu}\equiv (g-2)_{\mu}/2$ improve 
 constraints on  new physics. 
Using two types of 
 estimations of  $a_{\mu}^{had}$
(Davier \& H\"{o}cker (case A) and Jegerlehner2000 (case B)) 
we evaluate 95\% CL intervals  for a new contribution,
which can  be used to constrain any model beyond the Standard Model.
We apply  these intervals  to the general  2HDM ("Model II"), 
where up to now  one light neutral Higgs particle, $h$ or $A$,  
was allowed by the data.  
Assuming that only one Higgs boson contributes to 
$a_{\mu}^{2HDM}$  the two-loop calculation based on the case A 
 leads  to the exclusion 
of  scalar $h$ while a pseudoscalar 
 with  mass  between 10 and 70 GeV  ( for 
$\tan \beta \geq $ 20 ) is allowed. For case B
  the    upper limits for $\tan \beta$ for 
both scalar and pseudoscalar are obtained.
\end{abstract}
\pacs{PACS number(s): 12.60.Fr, 14.80.Cp}
\maketitle
\section{Introduction}
The precision measurement of $g-2$ for the muon is expected to shed 
light on  "new physics". The new E821 result based on 1999 data
\cite{Brown:2001mg} leads to a current mean of experimental results  
for $(g-2)_{\mu}$ (from \cite{Brown:2001mg})
$$a_{\mu}^{exp}\equiv{{(g-2)_{\mu}^{exp}}\over{2}}=11~659~203~(15)
\cdot 10^{-10},$$
where the  accuracy of this result (in parentheses) approaches the 
size of electro-weak contribution, 
$a_{\mu}^{EW}$.
The ultimate accuracy  of the E821 experiment is    4 $\cdot 10^{-10}$.

The QED and EW contributions to $a_{\mu}^{SM}$ are well under control.
The   predictions for the hadronic contribution $a_{\mu}^{had}$,
 which is  
more than forty times larger than $a_{\mu}^{EW}$, 
differ considerably among themselves both for  the  central value 
 and its uncertainty.
This uncertainty is presently  of order $ (7 - 10)  \cdot 10 ^{-10}$,
with the dominant  error coming  from 
  the $\alpha^2$ vacuum polarization contribution. 
Useful discussion of various estimations of $a_{\mu}^{had}$ 
 can be found in \cite{Czarnecki:2001pv}, see  also  more recent papers   
\cite{Yndurain:2001qw,Marciano:2001qq,nar2001,jeg2001,mel,DeTroconiz:2001wt,
Prades:2001zv}.

The difference between the experimental data, $a_{\mu}^{exp}$,
 and the Standard Model (SM) 
prediction [2-47], $a_{\mu}^{SM}$, defines the room for  "new physics".
Obviously the uncertainties of the hadronic contributions  influence  
 the estimation of a size of  new effects.
To illustrate  the present situation  we  calculate  
   95\% CL  intervals ($lim (95\%)$) for an 
allowed new contribution, $\delta a_{\mu}$,  
using  two representative SM predictions \cite{Czarnecki:2001pv}:  
one based on calculation
of $a_{\mu}^{had}$ with the  $\alpha^2$ vacuum polarization estimation
by Davier and H\"ocker \cite{Davier:1998si,davier} 
and the  other by   Jegerlehner \cite{Eidelman:1995ny,jeg2000}. 
We derive also   intervals  which may be relevant for models 
leading to  a  positive contribution only 
({\sl lim}$_{+}(95\%)$).
 
The obtained intervals we apply  to constrain the parameters of the 
 CP conserving 2HDM ('Model II') \cite{hunter,Ginzburg:2001ss}. 
This model is  based on the two doublets of 
complex scalar fields, and predicts existence of five Higgs particles: 
two neutral Higgs scalars $h$ and $H$, one neutral pseudoscalar $A$, 
and a pair of charged Higgses $H^{\pm}$. 
In MSSM, which  has a   Model II Higgs sector,
the mass limits (95\%CL)  from LEP experiments are, 
\eg~ for a maximal M$_h$ scan: 
$M_h (M_A)$  greater than 90.0 (91.9)  GeV 
and $\tan \beta<0.5$ or $\tan \beta>2.4$ 
\cite{MSSM}. The present  95 \% CL  limit on the SM Higgs 
mass is: $M_h >114.1$ GeV \cite{MSSM}. The SM Higgs particle with 
mass around 115 GeV (corresponding to a 
maximum likelihood ratio from LEP data \cite{MSSM})
 contributes 
to the  $a_{\mu}^{SM}$ at the level of  10$^{-11}$ (two-loop result
\cite{Czarnecki:1996sz}).

In the  non-supersymmetric 2HDM (II), which we study here, one light neutral 
Higgs boson $h$ or $A$ with mass below 50 GeV is still allowed by data [50-77],
 see \eg~ the LEP  
results from    Higgs-strahlung and  $hA$ pair production  
 \cite{Abbiendi:2001ug,abreu},
 from the Yukawa processes $Z\ra f \bar f h(A)$ with $f=\tau ~{\rm or }~b$ 
 \cite{bib-Ayuk,bib-Dyuk} and  the process $Z\ra h(A) 
\gamma$ \cite{Krawczyk:1999kk}, see also \cite{OPAL-new,Yu}. 
The dedicated fit to the EW precision LEP data 
performed within this model, even with masses of $h$ or $A$ below 20 GeV, 
is equally good as the corresponding fit in the SM or  MSSM 
 \cite{Chankowski:1999ta}. 
Unfortunately a potential of the HERA collider, 
discussed in \cite{Bawa:1995fp,Krawczyk:1996ed}, has  never been 
explored to put limits on very light Higgs bosons of the 2HDM (II).
Future Linear Collider experiments   are not expected either to close 
a light Higgs window
in the 2HDM (II) \cite{Choudhury:1997vc}, even these planned to run with 
a very high luminosity, once more at the Z-peak (GigaZ) 
\cite{Krawczyk:2001kf}. On the other hand one should keep in mind that some
theoretical arguments  disfavor   
the  Higgs scalar $h$ with mass below 90 GeV in 2HDM(II)
\cite{Kanemura:1999xf}.

A light Higgs 'could conceivable evade discovery at LEP and yet show 
up in a analysis of a low energy data' as $(g-2)_{\mu}$, 
as it was pointed out in paper \cite{Carlson:1988dp}. 
We have used previous experimental data and the SM prediction(s)   
  to constrain 2HDM (II) in \cite{Krawczyk:1997sm}. That one-loop 
analysis   led   to  a very small improvement in comparison to LEP limits.
It was expected  that with increasing precision 
the  $(g-2)_{\mu}$ measurements would  lead to more stringent constraints. 
As we have pointed out in \cite{Krawczyk:1997sm}, 
the sign of the one-loop  2HDM (II) contributions  to $a_{\mu}$
is correlated with  the type of the lightest particle: $h,A$ or $H^{\pm}$. We
  have described the condition which would lead 
to the exclusion at 95 \% CL  of a light $h$ or $A$. This condition is    
presently   fulfilled for a pseudoscalar
if the DH  result for $a_{\mu}^{had}$ is used in the one-loop approach, 
see also\cite{Dedes:2001nx,Larios:2001ma}. 
  However, as it was pointed out in 
\cite{Chang:2001ii,Cheung:2001hz}, the two-loop     calculation leads to  
 very different results.

In this study  we  constrain 2HDM (II) by the  new data \cite{Brown:2001mg} 
using  the two-loop calculation. 
We apply   two  different  hadronic contributions $a_{\mu}^{had}$
as mentioned above (DH=case A and J2000=case B) and   
two types of   
 95\% CL intervals ($lim$ and $lim_{+}$). Obtained constraints we combine 
with  constraints from other processes.

The one-loop analysis of the 2HDM(II) with results 
which partly overlap with results of this paper
can be found in \cite{Dedes:2001nx}, see also \cite{Larios:2001ma}.
 The two-loop analysis,
which we follow here, is given in   \cite{Chang:2001ii,Cheung:2001hz}.
Related study can be found in \cite{Wu:2001vq}.
Studies of $g-2$ for muon 
within the context of supersymmetric models
were performed  earlier \cite{MSSM-old1,MSSM-old2}, many new analyses 
have appeared recently 
\cite{MSSM-new,Chen:2001kn,Arhrib:2001xx,fl-MSSM}.
Dark matter problem in supersymmetric models 
is  discussed in \cite{dark}. Relations to neutrino masses and mixing
are studied within  supersymmetric and non-supersymmetric models
 in \cite{neutrino}.  
Other new  analyses made recently are \cite{newn}, analyses
 based on the general 2HDM (Model III) are   presented in 
\cite{modelIII,Wu:2001vq}.
Analyses of some aspects of cosmic rays can be found  in \cite{cosmic}.
Relations  between the electric and magnetic dipole moments
are studied in \cite{edm}.
An model-independent analysis was done in \cite{Einhorn:2001mf}. 

In Sec.II we discuss the new $(g-2)_{\mu}$ data and derive  limits for a 
new contribution
which can be used in any model beyond the SM. In Sec. III we apply 
the obtained limits to 2HDM (II) and derive  constraints on the
parameters of the model. Next we combine these constraints with
other experimental information. We study separately the one-loop and two-loop
results. Sec. IV contains  conclusions.  
\section{New g-2 data for the muon}
\subsection{New $(g-2)_{\mu}$ results} 
The current world average experimental data  on $(g-2)$ for muon 
averaged over the sign of the muon electric charge is given by (
from \cite{Czarnecki:2001pv}):
$$a_{\mu}^{exp}=116 ~592~023~\cdot 10^{-11}, \hspace {0.4cm}{\rm with}~ 
\sigma_{exp}=151~\cdot 10^{-11} .$$

The Standard Model prediction for this quantity 
consists of the QED, hadronic and EW contributions:
$$a_{\mu}^{SM}=a_{\mu}^{QED}+a_{\mu}^{had}+a_{\mu}^{EW}.$$
The  QED results can be found in  \cite{qed,ki}. Hadronic contributions
were obtained in: the leading vacuum polarization term (v.p.1)   
\cite{nar-old,Barkov:1985ac,Kinoshita:1985it,km,mar,re,Eidelman:1995ny,jeg2000,Alemany:1998tn,Davier:1998si,davier,Adel:1995bf,wort,nar2001,jeg2001,DeTroconiz:2001wt} and \cite{Erler:2001nx,Cirigliano:2001er},
 the higher order vacuum polarization term 
(v.p.2)
 \cite{nar-old,Kinoshita:1985it,Krause:1997rf,Alemany:1998tn}, 
the light-on-light term (lbl)
\cite{re,hayakawa,ll,Hayakawa:1998rq,nar2001,DeTroconiz:2001wt,Prades:2001zv,
lbl-new}. 
The EW results are given in 
\cite{EW-old,hhh,lu,Marciano:1988vm,Marciano:1992pr,Kukhto:1992qv,
Czarnecki:1995wq,Czarnecki:1996sz,Peris:1995bb,Degrassi:1998es}. 
A useful compilation of recent results 
is given in  
 \cite{Czarnecki:2001pv}, see also \cite{mohr,Studenikin:1999dw}.

The error of the hadronic contribution  dominates the
total error of  the SM predictions. Moreover the   hadronic 
contributions calculated by various authors as discussed in
\cite{Czarnecki:2001pv,Yndurain:2001qw,Marciano:2001qq,nar2001,jeg2001,
mel,DeTroconiz:2001wt,Prades:2001zv}
differ  strongly
leading to  distinct  estimations of a size of the SM contribution 
and therefore also of the 'beyond SM' effect.
All of these estimations 
 can  be divided  into two classes, depending on whether the 
SM prediction for $(g-2)_{\mu}$ 
is or is  not in agreement with the data. 
We think that it is sensible to study   separately consequences 
of these two  classes of the SM predictions.
It   allows to illustrate the present sensitivity to $a_{\mu}^{had}$, 
(see  \cite{Krawczyk:1997sm} for  previous results based on a similar approach,)
 moreover it may  serve as a guide for the future results, which 
obviously will belong to one of these two classes.

 In the analysis we consider two representative cases:  case A based on 
Davier and H\"ocker calculation of the leading 
vacuum polarization diagram (v.p.1) 
\cite{Davier:1998si,davier} 
and  case B based on the corresponding Jegerlehner 
calculation \cite{Eidelman:1995ny,jeg2000}, with a smaller and larger 
hadronic contribution (and its uncertainty), respectively.
In the DH analysis the $e^+e^-$ and $\tau$-decay data are used, while J2000
uses only the $e^+e^-$ data.
The higher order hadronic (v.p.2) contribution and the light-on-light 
contribution are taken from \cite{Krause:1997rf,Kinoshita:1985it} 
and \cite{ll,Hayakawa:1998rq}, respectively.    
 In the table below we collect, following \cite{Czarnecki:2001pv}, 
the corresponding SM contributions (and their uncertainties).

$$
\begin{array}{lrr}
case  &~{ \rm {A}}~[{ \rm in}~ 10^{-11}]&~{\rm {B~[in}}~ 10^{-11}] \\  
\hline
{\rm QED}   &~~~116~584~706 ~~~~(3)
  &~~~116~584~706~ ~~~~(3) \\
{\rm had}   & 6~739~~(67) &  6~803 ~~(114)  \\
{\rm EW}    & 152 ~~~~(4)  &  152 ~~~~~(4)   \\
\hline
{\rm tot}   &116~59 1 ~597 ~~(67)   & 116 ~591 ~661 ~~(114)
\end{array}   
$$

\vspace{0.3cm}

The difference between two  predictions of the SM,
case A and  case B, is 64 $\cdot $ 10$^{-11}$, 
with  the corresponding difference in accuracies equal to 
47  $\cdot $ 10$^{-11}$. It is worthwhile to  compare these numbers
to the ultimate accuracy of the E821 experiment 40 $\cdot $ 10$^{-11}$.

Results based on  other  estimations of the hadronic contribution, 
 can be easily obtained from results for 
considered cases, see discussion below.

\subsection{The room for  new physics}
Here we present results of  calculation of
  the   difference between the experimental and theoretical 
SM  results for $(g-2)_{\mu}$, which can be used for any model
 going beyond the SM.
 First we derive $\Delta a_{\mu}$, equal to the difference of the central 
values, $a_{\mu}^{exp}-a_{\mu}^{SM}\equiv \Delta  a_{\mu}$,
and  the  error $\sigma$ for this quantity. 
Knowing  $\Delta a_{\mu}$ and $\sigma$ one can calculate an allowed,
 at chosen confidence level (CL), interval
of an additional contribution. For the purpose of this study 
it is enough 
 to obtain $\sigma$ 
by adding in 
quadrature the corresponding experimental and theoretical errors
($\sigma_{exp}$ and $\sigma_{tot}$) and  
to assume a Gaussian distribution.    
 Under these assumptions we    calculate in both cases, A and B, 
the  $\delta a_{\mu}$ regions, symmetric  around $\Delta a_{\mu}$,
 allowed at 95\% CL.
$$
\begin{array}{lrr}
case  &~{\rm {A ~[in}} ~10^{-11}]
               &~{\rm {B ~[in}}~10^{-11}] \\  
\hline
\Delta  a_{\mu}(\sigma)    &426 (165) &~362 (189)  \\
\hline
{ lim(95\%)} &102\le\delta a_{\mu} \le 750 
     &~-8.65\le\delta a_{\mu} \le 733 \\      
\hline
{\rm positive~~} \delta a_{\mu}& 99.5\% &97.2\% \\
\hline
{ lim_{+}(95\%)} &109\le\delta a_{\mu} \le 744&
~~~~~28.5\le\delta a_{\mu} \le 696         
\end{array} 
$$
  In the  standard approach 
one obtains  intervals called {\sl lim}(95\%), 
see   the above table for results.
We see that although at the one sigma level,\ie ~for the interval 
$\Delta a_{\mu} \pm \sigma$, the allowed additional contribution to $a_{\mu}$ 
in cases A and B 
are  of a positive sign only, at the 2 $\sigma$ level or 95 \%CL
the more conservative estimation of the hadronic contribution to 
$a_{\mu}^{SM}$ (case B) makes  the negative $\delta a_{\mu}$ possible.
 The SM prediction lies within the 95\% CL interval 
for case B, while for case A it is outside the corresponding interval.

This difference leads to  different forms of the limits for  
a new contribution. The 95 \%CL interval
leads in case  A to  an {\sl allowed   positive} contribution 
({\sl an allowed  band})
 and at the same time to  the {\sl exclusion} of the negative contribution 
(at the higher CL level, see below). 
For the case B,  the positive  (negative) contribution  is only  
bounded from above (below) ({\sl   upper limits} for the absolute 
value of the new contribution).
That means that presently
 the accuracy of the theoretical predictions  for  the hadronic 
contribution (both $a_{\mu}^{had}$ and $\sigma_{had}$ matter !)
influences in qualitative way the constraints on the new physics
\footnote{In our  analysis from 1996 \cite{Krawczyk:1997sm} the 
corresponding differences  between the case A and B had only a
 quantitative character.}.

Now we discuss  consequences of the  present $(g-2)_{\mu}$ data for  
 models which can give contribution of  only {\sl  {one}} sign.
We see (the table above) that  for both A and B cases 
a negative $\delta a_{\mu}$ 
contribution is very unlikely: a positive (negative) contribution 
corresponds to 99.5 (0.5) \% CL for A, while for case B to  97.2 (2.8) \% CL. 
At this level models leading to {\sl only negative} $\delta a_{\mu}$ can be 
{\sl excluded} or saying differently,  within models 
 which give a definite  sign contribution to $a_{\mu}$
only these which give a {\sl positive} $\delta a_{\mu}$
can  be {\sl realized}  at 95\% (or higher) CL.

To obtain the allowed range of parameters
of such a  model we calculate the corresponding 95\% CL
intervals normalized to the positive contributions only \cite{data}.
Results for these intervals, called 
$lim_+(95\%)$,
are presented in the last row of the table above
\footnote{We found it difficult to apply the $lim_-(95\%)$
method to the negative contribution, which is  outside (or 
almost outside in case B) the 95\% region around $\Delta a_{\mu}$.}. 
Of course, these two types of   95 \% CL intervals,
{\sl lim} or {\sl lim$_+$}, 
will lead to very similar constraints  of parameters of the model
for  case  A,  since they will be obtained  
from only slightly different,  due to the 0.5\% change in normalization,
 allowed  $\delta a_{\mu}$. 
However,   significantly different constraints 
will arise for case B. Instead of upper limits 
obtained in the  {\sl lim}(95\%) approach, 
in the {\sl lim}$_+(95\%)$ method   an {\sl allowed  band} for a 
positive contribution is obtained.

We observe that the maximal  positive $\delta a_{\mu}$
values differ less  than 10 \% for all discussed cases (696 to 750 in 
10$^{-11}$). 
The minimal  positive $\delta a_{\mu}$ obtained  using $lim_{+}(95\%)$ method 
differ by a factor 3 in cases A and B, 109 and 28.5 in 10$^{-11}$, 
respectively.

Other recent estimations of the hadronic contributions  lead to
the following $\Delta a_{\mu}(\sigma)$ and intervals $\delta a_{\mu}$
in [10$^{-11}$]:
$$
\begin{array}{llcc}
Reference&&\Delta a_{\mu}(\sigma)&lim(95\%)\\
\hline
{ \rm Jegerlehner} [6] &{ \rm J2001}& 
376 (186)  &~12~ \leq \delta a_{\mu} \leq 740 \\
 { \rm Narison}\,\,\,\,\,\,\,\,\,\,[5] &{ \rm N}&
375 (170) &  ~41.8 \leq \delta a_{ \mu} \leq708\\
{ \rm Melnikov}\,\, \,\,\,\,[7]&{ \rm M}&\! 377 (216) &
  -47.2 \leq \delta a_{ \mu} \leq 801 \\
&&&\\
{ \rm DeTroconiz}&&&\\
{ \rm -Yndurain} [8] &{ \rm TY1}&363(184) 
&~2.52\leq \delta a_{ \mu} \leq 723\\
 & {\rm TY2}  &338(171)  &~3.08 \leq \delta a_{\mu} \leq 673\\
{\rm Prades} \,\,\,\,\,\,\,\,\,\,\,[9]&{ \rm P}&403(169)&~71.8\leq 
\delta a_{ \mu} \leq 734\\
\end{array}
$$
Note  similarities in the obtained $\Delta a_{\mu}$ values and  
at the same time  large differences in estimated uncertainties $\sigma$
for the first three analyses \footnote{for the M analysis 
we added errors in quadrature}.
The first two analyses are  similar to case A, 
while the third  one has properties of the case B.
 The maximal $\delta a_{\mu}$ differ from the  corresponding numbers for 
cases A and B 
 within 10\% , the minimal $\delta a_{\mu}$ differ much more.

The results of the newer analyses also are presented  in the above table. The  
TY1 analysis  which corresponds to  a (v.p.1) calculation based on the (new)
  $e^+e^-$ data, and 
 TY2 where  both the $e^+e^-$ and $\tau$ decay data are included 
\cite{DeTroconiz:2001wt}, the 
P  results  are based on the weighted average (for v.p.1) of the averaged
 estimation of (J2001 and TY1)  and (DH and TY2)
\cite{Prades:2001zv}. 
Note that all these estimations of $a_{\mu}^{had}$ are  based on a 
Chiral Model for the light-by-light contribution \cite{ll,Hayakawa:1998rq}. 
The Quark Model gives 
very different results. The corresponding limits were derived
in \cite{DeTroconiz:2001wt}: the estimations TY3 
and TY4 (analogous to TY1 and TY2, $\ie$ without and with $\tau$ decay data)
 leading to 
 -177 $\leq \delta a_{\mu} \leq$ 547 and -177$\leq \delta a_{\mu} \leq$ 497 in 
[10$^{-11}$], respectively.

In the following analysis we will apply intervals obtained for case A and B, 
a simple rescaling allows to translate the final  constraints to  results 
relevant for any other present or future estimations of $\delta a_{\mu}$.

\section{Constraining the 2HDM (II)}
\subsection{A model}
In the non-supersymmetric CP conserving 2HDM  
 the Higgs sector  contains the two neutral scalars, $h$ and $H$, 
pseudoscalar $A$ and charged Higgs bosons $H^{\pm}$.
Beside their masses, three parameters: $\tan\beta$, which is
the ratio of the vacuum expectation values of the Higgs doublets $v_2/v_1$,
 the mixing angle in the neutral Higgs sector $\alpha$,
and in addition one more parameter,
\eg~ the Higgs self-coupling $g_{hH^+H^-}$, specify the model.

In the  Model~(II) implementation of the 2HDM, one doublet of fundamental
scalar fields couples to the $u$-type quarks, the other to the $d$-type
quarks and charged leptons (this way FCNC processes 
are avoided at the tree level) \cite{hunter}. The ratios, relative to
the SM values, of the direct coupling constants of the Higgs boson
$h$ or $H$ to the massive gauge bosons $V=W$ or $Z$, and  to fermions 
{\it (\ie ~Yukawa couplings)}
can be determined via angles $\alpha$ and $\beta$
\cite{hunter,Ginzburg:2001ss}. For $\chi_i^h \equiv 
{g_i^h}/{(g_i^h)_{SM}}$ (and similarly for $H$) we have,
  in form suitable for discussion  simultaneously of $h$ and $H$:
\begin{eqnarray}
\chi_V^h=\sin(\beta-\alpha)\;\;,\;& \chi_V^H=\cos(\beta-\alpha),\;  \\
\chi_u^h=\chi_V^h+\cot\beta\chi_V^H,\;\;
&\chi_u^H=\chi_V^H-\cot\beta\chi_V^h, \\
\chi_d^h =\chi_V^h-\tan\beta\chi_V^H,\;\; &
\chi_d^H =\chi_V^H+\tan\beta\chi_V^h, 
\label{2hdmcoup-h}
\end{eqnarray}
with $(\chi_V^h)^2+(\chi_V^H)^2=1$\cite{Ginzburg:2001ss}.
A very useful {\sl pattern relation} among these couplings
holds for both $h$ and $H$ \cite{Ginzburg:2001ss}:
\begin{equation}
(\chi_u+\chi_d)\chi_V=1+\chi_u\chi_d.
\end{equation}
For $\chi_V^h=1$ all couplings of $h$ have the SM values, 
 couplings of $H$ to  gauge bosons are equal to zero while 
  couplings of $H$ to fermions may differ considerably 
from the SM values, for  small or large $\tan \beta$ 
\footnote{even for $\tan \beta=1$
they may differ by sign from the SM value}, 
\footnote{The perturbativity arguments constrain the $\tan \beta$ range 
between $\sim$ 0.2 and 200-300 \cite{limit}.}. 
For $\chi_V^H=1$  the  $H$ is   SM-like while 
$h$ has different properties (1-3), $\eg$ $\chi_d^h$ can be very large 
for large $\tan \beta$.
From (2) and (3) or (4) it  follows that for $\chi_V=0$ one obtains    
$\chi_u\chi_d=-1$.

For the pseudoscalar there is no coupling to $W/Z$.
The Yukawa couplings to fermions $\chi_d^A(\chi_u^A)$
contain  $\tan\beta$ ($\cot\beta$) factor. The $\chi_d^A$ 
is large  (small) 
for large (small) $\tan \beta$ value, respectively,
with the  opposite pattern  for $\chi_u^A$.

In this analysis the  Yukawa coupling  $\chi_d$, relevant for a 
Higgs boson coupling to a muon, plays a basic role.
It  is equal to $\tan \beta$ for a pseudoscalar and $H^+$ 
and, if in addition $\chi_V=\sin (\beta-\alpha)$=0, also for a scalar
(more precisely $\chi_d^h=\pm \tan \beta$).

At the two-loop level there appears a possibility of having a charged 
Higgs boson in a loop (see below). A coupling of $H^+$ to a scalar $h$ has 
 a form:
\be
\chi_{H^+}^h=(1-\frac{M_h^2}{2M_{H^+}^2})\chi_V^h+\frac{M_h^2-\mu^2}
{2M_{H^+}^2}(\chi_d^h+\chi_u^h),
\ee 
with the  normalization as for an elementary charged scalar particle 
in the SM. For  $\chi_V^h=0$ one gets
\be
\chi_{H^+}^h=\frac{M_h^2-\mu^2}
{2M_{H^+}^2}(\chi_d^h+\chi_u^h)= \frac{M_h^2-\mu^2}
{2M_{H^+}^2}\chi_d^h (1-(\frac{1}{\chi_d^h})^2).
\ee
\subsection{Existing constraints}
Main constraints of the parameters of the  2HDM (II) 
 come from LEP experiments, see also a discussion in \cite{where}.
From the $Z\ra Z h$ process upper limits on
  $\sin^2(\beta-\alpha)$ were derived 
\cite{Abbiendi:2001ug,abreu,bib-sine2LEP}.
From the tightest limits  \cite{bib-sine2LEP} it follows that 
 $\sin^2(\beta-\alpha)$ should  be 
smaller than 0.1 for the $0\mr  M_h\mr$ 50 GeV,
and  even below 0.01 for a lighter scalar. 
The data   for the cross section for the  pair ($h,A$) production,  
 proportional to $\cos^2(\beta-\alpha)$,  when combined with the $Z\ra Zh$ 
data leads to an exclusion of a small mass region in the ($M_h, M_A$) 
plane\cite{Abbiendi:2001ug,abreu}.
According to these data the 2HDM (II) may  accommodate a very light 
($\lsim 45 \gev$) neutral scalar $h$, with small $\sin(\beta-\alpha)$, 
 {\sl {or}} a very light pseudoscalar $A$ as long as:
$M_h+M_A \gsim 50$~\cite{Abbiendi:2001ug,abreu}, see also  
the newest results \cite{OPAL-new}.

 A neutral  Higgs particle at LEP I has also been searched for 
in the Yukawa process, 
$e^+e^- \rightarrow f {\bar f} h(A)$, where $f$ means here
{\it b} quark or $\tau$ lepton.
For a light scalar this  is  an additional,
and if  $\beta=\alpha$, the most important source of information.
A (still preliminary) ALEPH analysis of the Yukawa process  
  for a pseudoscalar \cite{bib-Ayuk} led to 
 the exclusion  at  95\% CL for  the ($\tb,M_A$) plane, 
allowing for a large $\tan \beta $, above 20,
  for  mass larger than 2 GeV. 
Similar   analysis was  performed by DELPHI for the 
  $b$ quark-couplings to   a scalar $h$ and pseudoscalar $A$. 
New measurements of Yukawa process at LEP by OPAL and DELPHI groups \cite{Yu},
 improve slightly these results, see discussion in Sec. IV.

A   measurements of $Z\ra h(A) \gamma $ performed by all experimental groups 
at LEP I was  used to 
obtain  upper limits (however weaker than from the Yukawa processes) and 
 lower limits on the Yukawa couplings  $\chi_d$ \cite{Krawczyk:1999kk}. 
Still  large part of the parameter space remains  unconstrained. 

Also the 
 dedicated   global fit to the EW precision data  in the 2HDM (II)
framework  allows for an
 existence of very light scalar or pseudoscalar 
\cite{Chankowski:1999ta}, for a partly constrained the heavy Higgs boson 
sector, including the $H$ and $H^{\pm}$ bosons.
Note that  the lower mass limit of $H^{\pm}$  estimated from the 
direct search at LEP is  78.6 GeV \cite{MSSM}.
The $b\ra s \gamma$ data interpreted in the 2HDM (II) give,
according to the newest results \cite{bib-bsg}, $M_{H^{\pm}}> 320$ GeV or even higher 
\cite{misiak2001}.
One of the important message from the global fit based on the EW precision 
data \cite{Chankowski:1999ta} is that for a light $h$ and large $\tan \beta$, 
 and for mass of $M_H$ below 1 TeV, an upper limit for $M_{H^{\pm}}$ 
 can be derived. In order to  agree with  the above lower mass limit 
based on $b\ra s \gamma$ data,  
  $\tan \beta $ should be smaller than 22 (28)
for a light $h$ with  mass 10(20) GeV. For the upper mass of $H$ equal to 
5 TeV, these maximal value of $\tan \beta$ increases by $\sim 3$.

In light of the above results 
one can conclude that  there is still a possibility of the existence of one
light neutral Higgs particle $h$ or $A$ with mass even below $\sim$ 40--50 GeV.
 Since  for a very light $h$  the  limit $\sin^2(\beta-\alpha) \ll 1$
 should  hold, the second scalar, $H$, 
is expected to mimic the SM Higgs boson couplings, 
as discussed in Sec.III A. Therefore it is reasonable to expect 
that its  mass  is equal
 to  $\sim$ 115 GeV (or slightly higher).

Other   low energy measurements do  not  contradict 
such scenarios.
The $\eta $ 
decay data \cite{pich} exclude   only  a very light $h$, 
with mass below 280 MeV.
The Wilczek process, $\Upsilon \ra h(A) \gamma$ \cite{Wilczek:1977zn}
\footnote{also a decay
$J/\psi \ra h(A) \gamma$ \cite{wil}}, has been measured  
by few groups \cite{wil,keh,narain,lee}. 
 Unfortunately the corresponding predictions  have large theoretical 
uncertainties   both due  to  the QCD and relativistic corrections, 
see \cite{hunter,wil-cor}. In addition, in some experimental 
analyses the production of the Higgs boson was treated according to  
the general 2HDM
while for a decay of such Higgs boson the SM rates were assumed, 
\eg ~\cite{narain}. 
All of these measurements of the $\Upsilon$ decays 
give only  upper limits for a coupling of the  Higgs 
boson to $b$ quark. Although for mass below 10 GeV these limits seem to be 
 stronger than others mentioned above,  
large uncertainties of various sources 
make   difficult to use these limits on a similar footing as the LEP ones.
In the present analysis on $g-2$ data , in which as we will see below 
also  {\sl lower limits} for the Yukawa coupling appear,  
the data for the Wilczek process,  even with  large uncertainties,  
will  play an important role in   closing low mass (below 10 GeV) 
part of a parameter space for 2HDM (II). 
In this analysis we apply three different constraints
from the   $\Upsilon \ra h(A) \gamma$ process, from \cite{keh} 
(denoted in figures as K), \cite{narain} (N) and \cite{lee} (L). 

Unfortunately there are no limits  from the HERA collider  
 on very light Higgs bosons 
\cite{Bawa:1995fp,Krawczyk:1996ed}. There are
  important upper limits for  the Yukawa couplings $\chi_d$ 
for $h$ and $A$ from the TEVATRON data in the  large mass region.
They were obtained originally for the 
Higgs bosons in  MSSM \cite{TEV}:
for mass  say 70 (120) GeV  $\tan \beta$ should be above 34 (60) \cite{TEV}.
These limits, rescaled by a factor $\sqrt 2$, should hold also for $h$ and $A$
 in  context of the  2HDM (II).

In Fig. 1 we present the  existing 95\% CL limits for a Yukawa coupling
$\chi_d$ for $h$ (solid line) and $A$ (dashed line) in the 2HDM (II).
 Upper limits are    from the Yukawa process 
(data from ALEPH \cite{bib-Ayuk} for $A$ and DELPHI \cite{bib-Dyuk} 
 for $h$ and $A$).
Lower limits from the $Z\ra h(A) \gamma$ processes \cite{Krawczyk:1999kk} 
measured at LEP are  also shown. In addition the upper 90\% CL 
constraints from the  $\Upsilon$ decay from \cite{keh} (K) are shown,
 rescaled by a factor 2 to take into account the preference of 
the Higgs boson to decay into tau leptons (It  occurs in 2HDM (II)
for the considered mass range, for $\chi_d$ bigger than 2.).
The TEVATRON limits \cite{TEV} (with a factor $\sqrt 2$) are  displayed
 as well.

The one-loop analysis based  on the previous $g-2$ for the muon results 
\cite{Krawczyk:1997sm} 
led  to a slight improvement of the upper limits of $\chi_d$ 
as compared  to  the Yukawa processes, for $M_{h(A)} \leq $  2 GeV  
(not shown). 
The new data  improve constraints of the 2HDM(II) considerably,
 what will be  shown below.\\
\begin{figure}
\includegraphics{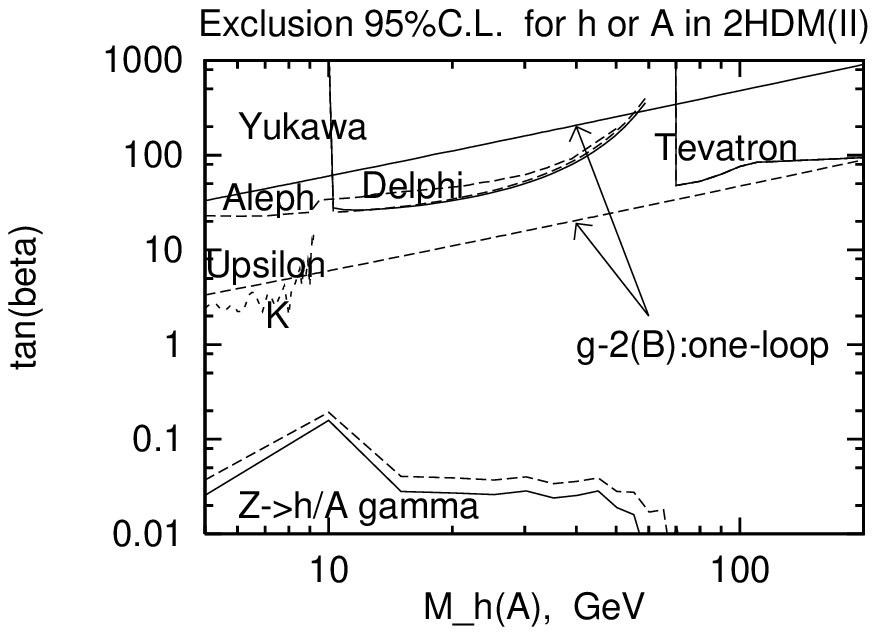}
\includegraphics{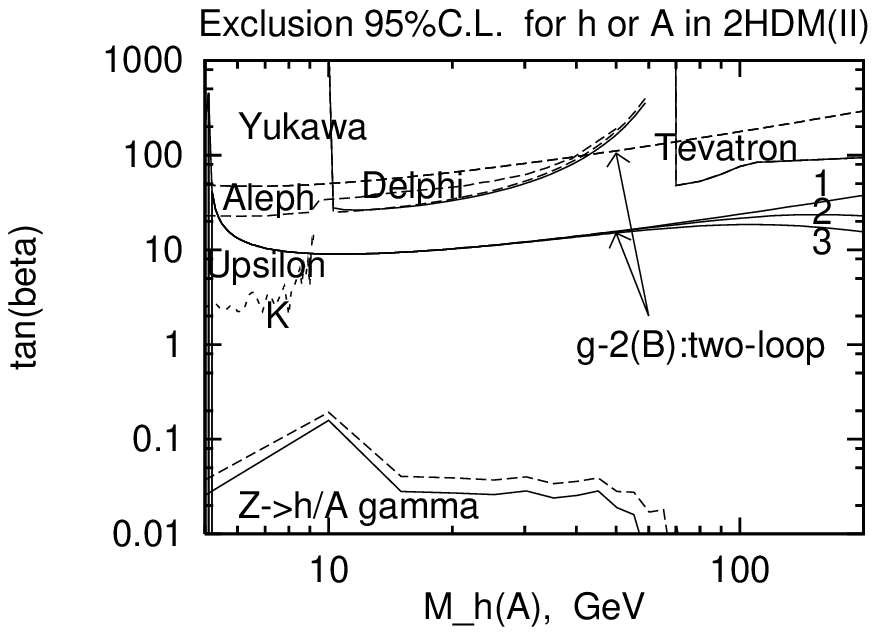}
\caption{\label{fig:fig1} The present upper and lower limits (95\% CL)
 for the Yukawa coupling 
$\chi_d$,  for a scalar $h$ ($\tan \beta$ if $\chi_V^h
=0$, solid line)  and pseudoscalar $A$ ($\tan \beta$, 
dashed line) as a function of mass. Upper limits are    
from the Yukawa process (ALEPH and DELPHI results)
and lower limits from the $Z\ra h(A) \gamma$ processes  
measured at LEP. In addition the upper 90\% CL constraints from 
 the  $\Upsilon$ decay (K), rescaled by a factor 2 are shown.
TEVATRON upper limits, both for $h$ and $A$, are presented.
The lines denoted ``$g-2(B)$'' corresponds to the upper limits for $h$ or $A$ 
obtained in this paper for case B  and  {\sl lim}(95\%):
 in the upper  panel  one-loop results, in the lower one 
the results of the two-loop analysis (for $\chi_V^h=0$),
  see text for details.}
\end{figure}

\subsection{2HDM(II) contribution   to $a_{\mu}$}
In the 2HDM (II) 
 neutral scalars $h$ and $H$, pseudoscalar $A$,  as well 
 the charged Higgs boson $H^{\pm}$ can contribute  to $a_{\mu}$.
There exist relevant calculations 
at the one-loop level \cite{hhh,lu,haber} 
 and the two-loop level  
\cite{Chang:2001ii,Cheung:2001hz,Arhrib:2001xx,hhhc}, see also  
 earlier  papers \cite{BW,Barr:1990vd}.
In contrast to  the one-loop approach where  each Higgs boson exchange
is given by  a separate diagram (Fig. 2),  
  various Higgs particles  may appear in the same  two-loop 
diagram, see Fig. 5(right).    

We  assume that the lightest Higgs boson, $h$ or  $A$, 
dominates the full 2HDM (II) contribution, $\ie$ $a_{\mu}^{2HDM} \approx 
a_{\mu}^h$, or $a_{\mu}^A$
 ({\it a simple approach}, see also \cite{Krawczyk:1997sm}). 
This approach should hold for  masses below 50 GeV, as discussed in Sec. 
III.B. For higher masses, which also  are   considered here,
this should be treated as an assumption of a large gap between 
Higgs bosons masses. Since the charged Higgs boson mass should be bigger 
than 320 GeV \cite{misiak2001},  we do not consider $H^+$ 
to be a lightest particle of  the model.

We calculate   
separately  the one-loop and two-loop  contributions to both 
 $a_{\mu}^{h}$ and  $a_{\mu}^{A}$.
This way the importance of the two-loop diagrams can be seen, 
 moreover our results can be easily compared with other 
one-loop calculation , $\eg~$ \cite{Dedes:2001nx}. 
For a pseudoscalar  $A$ the  two-loop contributions are  due to 
diagrams with fermion loops only. For a scalar 
the  fermionic, $W$ and 
a charged Higgs boson loops can contribute.
However, the $W$ and charged Higgs boson contributions 
are expected to be strongly suppressed for a small mass of $h$, 
where, according to the 
LEP data,  the  coupling $\chi_V^h$ should be small 
(see discussion  in Sec. III.B).
For a simplicity in calculating of the two-loop contribution
 for $h$ we explicitly 
assume for a whole mass range $\chi_V^h=\sin(\beta-\alpha)$=0, 
what means a domination of  the fermionic loops in   $a_{\mu}^{h}$.   
The more extensive consideration will be given elsewhere \cite{chkz}.

For  one- and two-loop approach we derive  constraints on Yukawa coupling
for $h$ and $A$
obtained from the estimated   $\delta a_{\mu}$ intervals  (Sec. II.B),
 by  taking $a^{2HDM}_{\mu} = \delta a_{\mu} $. 
\subsubsection{One-loop calculation}
The set of the relevant diagrams is presented in Fig. 2.
\begin{figure}[b]
\includegraphics{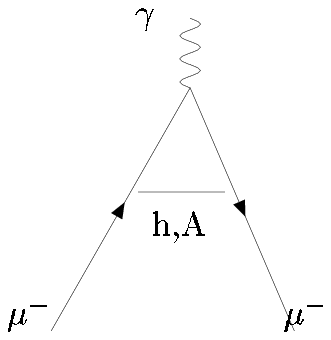}
\hspace*{0.9cm}
\includegraphics{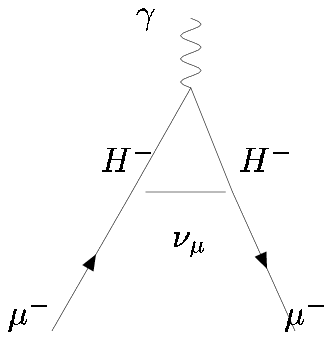}
\caption{\label{fig:fig2} 
{  One-loop contribution  to $g-2$ for muon due to a neutral 
scalar $h$ (or $H$), pseudoscalar $A$ and a charged Higgs boson $H^+$
 exchange. }}
\end{figure}


\paragraph{Individual contributions.}
The relevant one-loop formulae from the Appendix of 
paper \cite{Krawczyk:1997sm}, based on results 
\cite{hhh,haber,lu}, are given by 
(${\Lambda}=h,~A ~{\rm or} ~H^{\pm}$)
\begin{displaymath}
{ a_{\mu}^{\Lambda}|_{one-loop}}=
{\frac{f_{\Lambda}^{2}}{8{\pi}^{2}}} {{L}_{\Lambda}},
~~~~{f}_{\Lambda}\equiv{{g ~m_{\mu}}\over{2~ M_{W}}} ~\chi_d^{\Lambda}.
\end{displaymath}
If  $\beta=\alpha$, the coupling $\chi_d^{\Lambda}$ is universal for 
$h,A$ and $H^{\pm}$, and it is equal to $\tan \beta$ (see sec. III. A).

The integral ${L}_{h(A)}$  
for the neutral Higgs boson contribution is given by 
(with $z=({{m_{\mu} }/{M_{\Lambda}}})^{2}$):
\begin{displaymath} 
{{L}_{h(A)}}(z)=z {\int_{0}^{1}}{dx}{\frac{Q_{h(A)}(x)}{x^{2}z+(1-x)}}. 
\end{displaymath}   
with:${Q_{h}(x)}={{x^{2}}{(2-x)}}, ~~{Q_{A}(x)}=-x^{3}.$
The charged Higgs particle exchange is described
by: 
\begin{displaymath}
{\cal{L}_{\pm}}(z)=z{\int_{0}^{1}}{dx}{-{x}(1-x)\over{(x-1)z+1}}. 
\end{displaymath}
The scalar contribution $a_{\mu}^h$ \footnote
{For a larger mass it also can be used to represent  $H$. 
{Note, that the SM Higgs boson  contribution  (with mass $\sim$ 150 GeV) is  
included in the (two-loop) prediction  for $a_{\mu}^{SM}$ .} }
 is positive whereas the
pseudoscalar   and the  charged Higgs boson give negative contributions.
Each one-loop contribution  $a_{\mu}^{\Lambda}$ 
 disappears in the limit of large mass like 
$m_{\mu}^2/M_{\Lambda}^2 \ln (M_{\Lambda}^2/m_{\mu}^2 )$, 
see also \cite{Carlson:1988dp,Dedes:2001nx}. 
At small mass each  contribution reaches its maximum (or minimum if 
negative) value.

\begin{figure}[ht]
\includegraphics{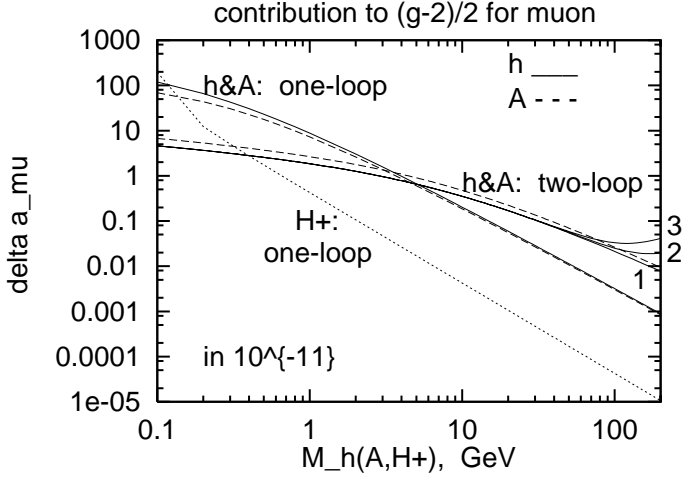}
\caption{\label{fig:fig3}  The (absolute value of)
individual contributions to $a_{\mu}$ from 
 a neutral scalar $h$ (solid line),  a pseudoscalar $A$ 
(dashed line) and  a charged Higgs boson $H^+$ (dotted line). For results 
based  on  one-loop calculation:   $A$ and $H^+$ contributions are 
 negative. Two-loop diagram contributions 
only for $A$ and $h$ (denoted ``1'') are
 based on the  down-type fermion loops. 
For $h$ also results with additional  charged Higgs boson loop are shown 
(coupling of $H^+$ equal to a first term in eq.6. with $\mu=0$) 
line ``2''(``3'') corresponds to $M_{H^+}$=800(400) 
GeV. The two-loop $h$ ($A$) contribution  is negative (positive). 
Yukawa couplings as in the SM are assumed.} 
\end{figure}

The individual (absolute value of) contributions, with  couplings as in the 
SM,\ie~with $\chi_d^{\Lambda}=1$ , are shown in Fig.3.
For mass above 0.2 GeV the charged Higgs contribution is much smaller 
than the contributions due to neutral Higgs bosons $h$ and $A$. 
One observes that  the $h$ 
and $A$ give practically the same  contribution for mass larger than few GeV,
  only  difference being in sign.   
In principle one should take into account a
 possible cancellation of these contributions, especially 
for masses above 50 GeV, see
 \cite{chkz}.

\begin{figure}[ht]
\includegraphics{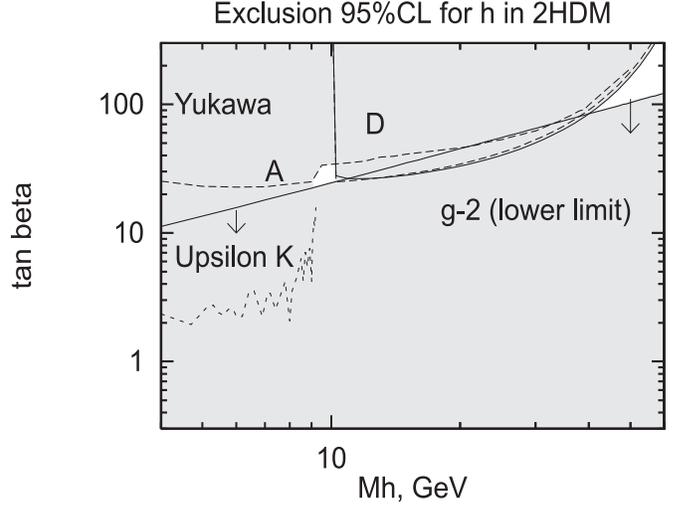}
\caption{\label{fig:fig4}
The exclusion plot for a Yukawa coupling $\chi_d^h$ ($\tan \beta$ for 
$\beta=\alpha$) for a scalar in the 2HDM (II) (a one-loop calculation).
The lower limit from the $(g-2)_{\mu}$ data for case A, 
the allowed region lays above the line with arrows.
Upper limits from the Yukawa process (A=ALEPH and D=DELPHI), 
and the $\Upsilon$ decay (K). White areas are  allowed at 95\% CL 
(both $lim$ and $lim_+$).}
\end{figure}

\paragraph{Constraints.}

First we present constraints   
based on the  standard   95\% CL intervals, {\sl lim}(95\%). 
For case B this  approach  leads to the {\sl upper limits}
on the Yukawa coupling $\chi_d$ for a pseudoscalar ($\tan \beta$)
and for a scalar ($\tan \beta$  if $\beta=\alpha$). 
Results are shown in  Fig. 1 (upper)  as lines denoted  ``$g-2$(B):one-loop''. 
For case A  one obtains an {\sl allowed band} for the Yukawa coupling 
of a scalar  only (only $h$ gives a positive contribution). 
It can be found in Fig. 4 (only lower edge is shown) 
and in Fig.6 (upper), as a  region between lines denoted A/B and A.
A pseudoscalar is excluded in this case.

Both for B and A case,  the 
 limits for $\chi_d$ rise with
 mass of the Higgs particle, what reflects the decrease of the corresponding 
$a_{\mu}^{h,A}$ terms with increasing  mass.
For case A, only large $\tan \beta$ (for $\beta=\alpha$) greater than 10
is  allowed  for mass of $h$ above 5 GeV, see Fig. 4.

Next we discuss results based on $lim_+(95\%)$ interval.
For case A one obtains  the {\sl allowed band}
for the Yukawa coupling  $\chi_d^h$, which practically  
overlap with the considered above  
{\sl lim}(95\%) band,  presented in  Figs.4 and  6(upper). Now, also
 for  case B  instead of the upper limits 
an {\sl allowed band} for the Yukawa coupling  appears for $h$ (only). 
In Fig. 6 (upper) these two allowed  bands for $h$ are compared:
a region between
lines  A/B and  A, and  a wider  region  between 
lines A/B  and B$_+$. 
For both cases, A and B, one  can exclude the negative contribution,
as they may be  realized at the level 0.5 \% or 2.8\%, respectively. 
This means an {\sl exclusion of a pseudoscalar}.

\subsubsection{Two-loop calculation}
The two-loop diagrams, see Fig. 5,  can   give  large contributions since
they allow to avoid one small Yukawa  coupling with muon  in favor of the 
coupling with other, potentially  heavy, particles circulating  in the loop 
 \cite{BW,Barr:1990vd,Chang:2001ii,Cheung:2001hz,Arhrib:2001xx}. 
In addition to the mass effect, 
such contribution can be enhanced further by an 
additional factors (eqs. 1-3,5).

In principle diagrams with a  $Z$ boson, instead of the exchanged photon,  
may appear. However they are expected to be small \cite{Chang:2001ii}
and are  neglected in this analysis.

\begin{figure}
\includegraphics{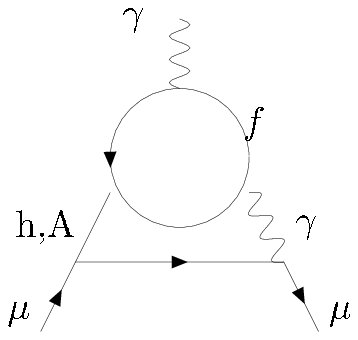}
\hspace*{0.5cm}
\includegraphics{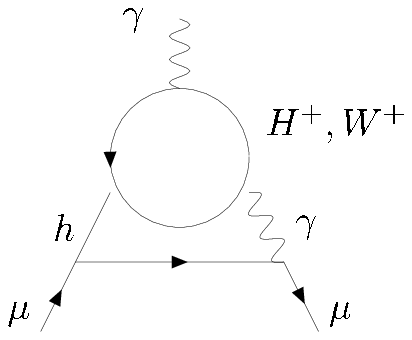}
\caption{\label{fig:fig5} 
  Two-loop contributions to $a_{\mu}$ from a light $h$ or $A$ with a
 fermionic loop (left); two-loop diagram for a light $h$ with
 a charged Higgs boson $H^+$ or $W^+$ loop (right).
}
\end{figure}

Below we consider the diagrams with fermionic loops  Fig. 5 (left), which are 
the only two-loop contributions for a pseudoscalar $A$. 
For  $h$  in addition also 
a charged Higgs boson loop (Fig. 5 (right)) is taken into account. 
In calculation of the two-loop contributions we take  $\chi_V^h$=0, 
as discussed above.
With this condition one can neglect  a $W$-loop contribution.

\paragraph{Individual contributions.}
The contributions from  diagrams with fermionic loops
 presented in Fig.5 (left)  
are given by the following formulae for $\Lambda=h,A$ \cite{Chang:2001ii,Cheung:2001hz,Arhrib:2001xx} 
\footnote{The two-loop contribution for a charged Higgs boson
 can be found in \cite{hhhc}.}:  
\begin{equation}
{ a_{\mu}^{\Lambda}|_{two-loop}}=
{\frac{ f_{\Lambda}^{2}}{8{\pi}^{2}}} {\frac{e^2}{\pi}} \xi \kappa  
{\tilde{L}_{\Lambda}^f},
~~{f}_{\Lambda}\equiv{{g ~m_{\mu}}\over{2~ M_{W}}} ~\chi_d^{\Lambda}.
\end{equation}
The $\kappa$ parameter is equal to 1 for a pseudoscalar $A$, for a scalar $h$
we have 
\begin{eqnarray}
\kappa={\frac{\chi_u^h}{\chi_d^h}} \;\;\; 
{\rm for} \; f="u-type \;\; {quarks",} \\ \nonumber
\hspace*{-2cm}\! \kappa = 1 \;\; {\rm for}~ f="d-type \;\; {quarks}" \\ \nonumber
\hspace {2cm}{ \rm for}~ f="charged \;\; leptons".
\end{eqnarray}
The $\xi$ parameter is equal to 1 for leptons and to $N_c Q_q^2$ 
for a quark $q$ with the  electric charge $Q_q e$, $N_c=3$.

The integral ${\tilde{L}_{h(A)}^f}$  
for the neutral Higgs boson contribution with a fermionic loop is given by:
\begin{equation} 
{\tilde{L}_{h(A)}^f}(z)={\frac{z}{2}}
{\int_{0}^{1}}{dx}{\frac{\tilde Q(x)_{h(A)}}{x(1-x)-z}}
{\ln{\frac{x(1-x)}{z}}},
\end{equation}   
$$z=( {\frac{m_f}{M_{h/A}}} )^2,$$ 
and $
{\tilde Q_{h}(x)}=-(1-2x(1-x)), \;\;{\tilde Q_{A}(x)}=1.$
A diagram with the $H^+$ loop presented in Fig.5 (right) contributes 
to $a_{\mu}^{h}$ with the integral 
$\tilde L_h^{H^+}(x)$  given by 
a similar expression   as for fermionic loop  (eq.11), with  
  $\tilde Q_h^{H^+}(x)=-(x(1-x))$ \cite{Arhrib:2001xx}. 
The corresponding  coefficients for 
its  contribution to $a_{\mu}^h$ 
are: $\xi=1$ and  $\kappa=\chi_{H^+}^h/\chi_d^h$.  
 
The integrals describing two-loop contribution with a fermionic  
loop is negative for $h$, 
while positive for $A$, on contrary to the one-loop results.
In Fig. 3  the (absolute value of) 
two-loop fermionic contributions, with  $\mu,\tau,b$ loops
for $h$ (denoted ``1'', solid line) and $A$ (dashed line), 
are presented for $\chi_d=1$. 
These two-loop  contributions 
dominate for masses above few GeV 
over the corresponding (absolute value of) one-loop terms for both 
$h$ and $A$. 
They have a milder than  the one-loop contribution dependence on 
the mass of $h$ (or $A$), namely   $\ln(m_{f}^2/M_h^2)$.
 From the figure one can read that 
the change of the sign of  the sum of the one- and 
two-loop terms appears at mass of $A$ ($h$) equal to 3 (5) GeV.

The $top$ contribution  dominates for $\chi_u=1$
in the  large mass region, being of order  $\sim 1 \cdot 10^{-11}$
 (not shown). 
However it  does not  play an important role in the present analysis,
since its contribution is proportional to $\chi_d\chi_u$ 
 equal to -1 for   the pseudoscalar $A$ and  (if  $\chi_V^h$=0) also  
for $h$.  So, the 
$top$-loop contribution  has no additional enhancement factor,
and for large $\chi_d$ the 
$d$-type fermion loops  dominate,   both in  
 $a_{\mu}^h$ and $a_{\mu}^A$.

The charged Higgs boson loop  contribution gives, 
for $\chi_V^h$=0 and  $\kappa=\frac{M_h^2-\mu^2}{2 M_{H^+}^2}>$ 0,
a negative contribution to $a_{\mu}^h$.  
This contribution (absolute value of) 
rises with a mass of $h$ like ${M_h^2}/{ M_{H^+}^2} 
(\ln M_{H^+}^2/M_h^2+5/3)$. 
Its effect can be seen   at  $M_h$
above 100 GeV. In Fig.3 these results for $a_{\mu}^h$ are presented for  
$\mu=0$
(lines ``2''for $M_{H^+}=$800 GeV and ``3'' (400 GeV)). 
It is clear that a charged Higgs boson loop with the  parameters 
as described above leads to a small 
modification at large mass of $h$ only.  
More detailed discussion will be given elsewhere \cite{chkz}.

\paragraph{Constraints.}
A full two-loop calculation, with both  one- and two-loop diagrams 
  included, 
 leads to results which differ significantly from 
the ones based on 
the  one-loop diagrams  only. The main difference is related to the 
fact that for masses
 above  5 (3) GeV, a  scalar (pseudoscalar) contribution to  $a_{\mu}$
has opposite  sign as compared to   the corresponding one-loop contribution.
It means that now for each scenario, with a light $h$ and a light  $A$,
 the contribution 
can be positive or negative  depending on mass. It is not the case
for  the one-loop terms where the corresponding 
 contributions have a fixed sign.  

In the derivation of the constraints  in  two-loop approach
we take also into account the non-leading terms, 
namely the $top$ contribution
and for  $h$ also a term $\sim \chi_u \chi_d$ due to the 
 charged Higgs boson loop 
(6). The results are as follows.

The case B allows in  the standard approach  for a negative and positive 
contributions to $a_{\mu}$. Therefore
the upper limits for $\chi_d$ 
exist for the whole mass ranges for both $h$ and $A$. In
Fig. 1 (lower ) we present
 the obtained limits (lines denoted ``g-2(B):two-loop'')
 for a  mass region above 5 GeV. Here 
the upper limit for pseudoscalar was obtained from the maximal positive
$\delta a_{\mu}$ value, while for a scalar -- from a maximal negative one,
see the first table in Sec. II.B.
This explains why a scalar is now constrained more tightly 
than a pseudoscalar,
contrary to limits based  on the one-loop calculation (Fig. 1 (upper)).
For a scalar results obtained with  only down-type 
fermions $\mu, \tau, b$  included in the two-loop calculation
are represented by  the  line denoted  ``1'' . Lines 
``2'' and ``3'' are obtained  if in addition one takes into account the 
charged Higgs boson-loop, as described above. 
Note that if one compares the two-loop constraints for a scalar 
to  the one-loop constraints for a pseudoscalar 
these for a scalar are more tight for   mass above few GeV, due to a 
weaker  mass dependence of the $a^h_{\mu}|_{two-loop}$, see Fig. 3.

The case A leads to an allowed band for a  positive contribution to $a_{\mu}$,
$\ie$ for a scalar $h$ with  mass below 5 GeV,
 and for a pseudoscalar $A$ with  mass above 3 GeV, similar results
were   found in \cite{Chang:2001ii,Cheung:2001hz}. 
Obtained constraints are presented   in Fig.6 (lower) and in Fig. 7 
as the regions between lines A/B and A. 
In the same figures also results for case B obtained with  $lim_+(95\%)$ 
are shown (regions between lines A/B and B$_+$).
It is not clear however whether this approach should be used here, 
as  for both,  a light $A$ and  a light $h$, scenarios,
   both positive and  negative 
contributions to $a_{\mu}$ are possible, as discussed in Sec. III. C.

\subsection{Combined 95\% CL constraints of the 2HDM (II) }
When the above constraints obtained from a new $g-2$ for the muon  measurement
are added to the existing constraints from other processes 
 discussed in Sec. III. B. (and  also  in \cite{Yu})   
interesting conclusions can be reached in the 2HDM (II) for  both  scenarios: 
with  a scalar $h$ and a pseudoscalar $A$ being the lightest particles in 
the model. We start discussion of the results based on
 one-loop calculation
which can be compared with similar analysis \cite{Dedes:2001nx},
then results of the two-loop analysis are presented.

The constraints are obtained for (absolute value of)
$\chi_d$, which is equivalent to $\tan \beta$ for pseudoscalar and  
for scalar, provided in the latter case $\chi_V^h$ is equal or close to zero. 
Only in the  calculation of the two-loop contribution we explicitly use the
assumption $\chi_V^h$= 0. For a simplicity of the discussion we will use
below the  $\tan \beta$ to represent  the Yukawa coupling $\chi_d^h$
 for a scalar in all cases.
\subsubsection{Allowed regions from one-loop results.}

The upper limits for $\tan \beta$ for a pseudoscalar $A$
which were  obtained from $g-2$ data for case B   
are much tighter than  the limits 
from other experiments for   mass above 10 GeV.
  Still  a window with a 
light $A$  is open for $\tan \beta$ below $\sim$ 10 (Fig. 1(upper)). 
For a scalar $h$  a weak improvement is observed only for  a mass range 
between 60-70 GeV where $\tan \beta $ has to be  lower than $\sim$ 300.  

The $(g-2)_{\mu}$ results for  case A  rule out a pseudoscalar, for a scalar  
 they improve considerably existing limits. The obtained  
 an allowed  band in the ($\tan \beta, M_h$) plane, is equivalent to
 existance of  both
the upper (as above) and   also the  lower limits for the Yukawa coupling.
 An allowed  by all experiments region  
 appears for a scalar $h$ with  mass  
between  40 and  70 GeV at  $\tan \beta$ above 80 (Fig. 6 (upper)). 
In addition  a  small region of  
 mass of $h$ around  10 GeV and  $\tan \beta$ between 25 and 35
remains allowed (Fig. 4), see also \cite{Dedes:2001nx} where the same result
was obtained.
Note, however, that this small allowed area  
can be closed by taking into account the  mentioned in Sec. III.B  
constraint from the global fit \cite{Chankowski:1999ta}. 
This constraint, for  the charged Higgs boson mass above 300 GeV, 
forbids   $\tan \beta$ to be larger than 22 for $M_h$=10 GeV 
\footnote{see Fig.9a from \cite{Chankowski:1999ta}}
\footnote{The newest data on the Yukawa process \cite{Yu}
 support this finding.}. 

 At the  one-loop level it  is sensible to study $lim_+(95\%)$ intervals, since 
two scenarios of the 2HDM(II), with a light $h$ or  $A$,
give contributions of a definite sign independently on mass.
Pseudoscalar, giving a negative contribution, is ruled out.
An allowed $lim_+(95\%)$ band for  case B  obtained for a scalar 
is wider than for discussed above case A
(compare  the region between lines A/B and B$_+$ with the region
 between A/B and A in Fig.6 (upper)).
Therefore also the allowed region in the ($\tan \beta,M_h$) plane
 is much larger for case B than for case A. It starts at mass 10 GeV
and stops at  70 GeV with $\tan \beta$ between 10 and ~$\sim$ 300.

In Fig. 6 (upper) we plot also an 
the expected upper 95 \% CL limits from process $gg\ra h \ra \tau \tau $ at 
the $ep$ collider HERA (dotted lines for lower and higher 
luminosity), from {\cite{Bawa:1995fp,Krawczyk:1996ed}}. 
These measurements may help to cover the 
low mass region for  $h$ and $A$ at the intermediate Yukawa coupling
$\tan \beta$. 

\subsubsection{Allowed regions from two-loop results.}
The two-loop diagrams   give  dominant
contributions to $a_{\mu}^{2HDM}$ 
for mass above 3 GeV (5 GeV) for a pseudoscalar (scalar)
and these contributions 
 have   reversed signs relative to the one-loop results.
Two-loop analysis for these regions  (based on a sum of the one- and 
two-loop diagram contributions)
leads to  similar conclusions as the one-loop  one,
 with  a reversed role of a scalar and  a pseudoscalar. In particular 
 now  the (standard 95\% CL)  upper limits obtained for case B  
 for scalar, for $M_h>$ 10 GeV,
are much tighter in comparison with   limits from other experiments
(Fig. 1 (lower)).
 Still a window with a 
light $h$  is open for $\tan \beta$ below 10. 
For a pseudoscalar the improvement is weak , being 
limited to a mass region between 40-70 GeV with $\tan \beta \le $  100.  

The constraints obtained in  case A  in a form of allowed bands
 improve considerably
the existing up to now  limits for both  scalar and pseudoscalar since 
in addition to the mentioned above 
 upper limits  also  the lower limits appear (Fig. 6 (lower) and Fig.7).
The allowed by $(g-2)_{\mu}$ data bands for a scalar $h$ is situated  
now below $M_h$=5 GeV. This region is  however  excluded by the $\Upsilon$
 decay data (Fig. 6 (lower)).
On the other hand a pseudoscalar with mass above 10 GeV 
and $\tan \beta$ larger than  20
is still in agreement with  existing data (see Fig.7). 
The TEVATRON data 
close practically the region of the mass above 70 GeV for a case A.

If for case B the  $lim_+$ interval is applied this  region of mass $M_A$ 
above 70 GeV  is still open.  

\subsubsection{Discussion}
As $(g-1)_{\mu}$ data favor a positive additional contribution
it is clear that  at one-loop level the most stringent limits  of the new $g-2$ data 
 for the muon  can be 
driven for a pseudoscalar, the  
 case A
leads  even to its  exclusion.

 The two-loop calculation leads to a radical  change of the picture.
The two-loop diagrams  give  the dominant
contributions for mass above 3 GeV (5 GeV) for a pseudoscalar (scalar)
and with a reversed sign as compared to the one-loop results.
Constraints are now tighter for a scalar.

 We point out a role in the present analysis 
of the low energy 
measurement of  the Wilczek process $\Upsilon \ra h/A \gamma$ in closing 
the window for mass below 10 GeV for both a scalar and pseudoscalar. 
The TEVATRON results close a part of the  large mass
and large $\tan \beta$ region.  The global fit adds an 
important constraints for $h$ with mass around 10 GeV.
New measurements of Yukawa process at LEP by OPAL and DELPHI \cite{Yu},
which were presented recently  do not change our qualitatively 
our conclusions, see also below.

\section{Conclusion}

We described  the room for a new effects as follows from the  recent
 $(g-2)_{\mu}$ measurements  and from new theoretical estimations of 
$a_{\mu}^{SM}$. 
For two  SM predictions arising from  different values of $a_{\mu}^{had}$,
case A (DH) and case B (J2000),
 we evaluate 95\% CL intervals for a new contribution.
They can  be used to constrain parameters of any model beyond 
the Standard Model.We show how  these  constraints   
depend on the size of the $a_{\mu}^{had}$ (case A and B),
and on the type of limits ($lim$ or $lim_+$).
The   {\it upper limits} for the negative and positive contributions, 
obtained if the  prediction B and the {\sl lim}(95\%) 
are used in the analysis, are to be contrasted with  
{\it the allowed band}, 
obtained in case A for a  positive contribution only.
For both A and B cases a negative $\delta a_{\mu}$ contribution 
is very unlikely: a positive (negative) contribution corresponds to 
99.5 (0.5) \% CL for A, while for case B to  97.2 (2.8) \% CL. At this 
level  models leading to only  negative $\delta a_{\mu}$ can be excluded, 
and vice versa models which lead to the positive contribution only 
should be accepted as   possible models. For such models the derived
 $lim_+(95\%)$ estimations differ from  those based on 
$lim(95\%)$ considerably for  case B.

We applied the obtained intervals 
 to  constrain  parameters of the 2HDM (II) using a simple approach,
where only one Higgs boson, $h$ or $A$, contributes.
In the one-loop calculation  
a light scalar scenario leads to the positive, whereas the
one with a light pseudoscalar to the negative contribution to $a_{\mu}$, 
independently of mass.
In the two-loop analysis, based on a sum of the one- and two-loop
diagram  contributions, the situation changes drastically.
Now the positive contribution can be ascribed
to a scalar $h$ with mass below 5 GeV or  a  pseudoscalar $A$ with
mass above 3 GeV. 
In both,  one- and two-loop,  approaches
 we derive tight constraints on the Yukawa couplings to
 muon of $h$ and $A$.
When these constraints are combined with 
 constraints 
arising from  other experiments, especially the Wilczek process and 
the Yukawa processes at LEP and TEVATRON,    large part of the 
parameter space for light $h$ or light  $A$ can be excluded.

Our results obtained for the case B lead to an improved upper limits 
for both $h$ and $A$. For 
a more constraining case A  our results including 
 all other existing constraints  are as follows
(in parenthesis the limits obtained if the  newest DELPHI data \cite{Yu}
are included ):

The one-loop calculation
excludes  a  pseudoscalar while allows for an existance of    
a light scalar $h$ 
with mass above 40 GeV (50 GeV) and below 70 GeV, and  $\tan 
\beta$ larger than 90.

The two-loop  analysis allows for an existance of 
a pseudoscalar with mass between $\sim$ 10 GeV 
(25 GeV)  and 70 GeV,
 and $\tan \beta$ above 20 (30). A light scalar
being excluded by combining the $(g-2)_{\mu}$ and  
the $\Upsilon$ decay data.

\section{Acknowledgments.}
I am  grateful to 
Jan \.Zochowski for his collaboration at the early stage of this work,
and to Piotr Zalewski for his important suggestions. 
I thank  Marek Szczekowski, Piotr Chankowski and Stefan Pokorski 
for helpful discussions on this analysis and Andrzej Czarnecki and William 
Marciano for clarification of the newest results for $g-2$ for muon.
I am grateful to Bohdan Grzadkowski and Wai-Tee Keung for a discussion 
on two-loop calculation and to Fred Jegerlehner for useful information.
I am indebted also to  T. and P-M.Krawczyk 
for their help in preparation of this paper. 
Supported in part by the Polish Committee for Scientific Research, Grants
5 P03B 121 20 i 2 P03B 05119,  and the European Commission 50th framework 
contract HPRN-CT-2000-00149.


\begin{figure*}[t]
\includegraphics{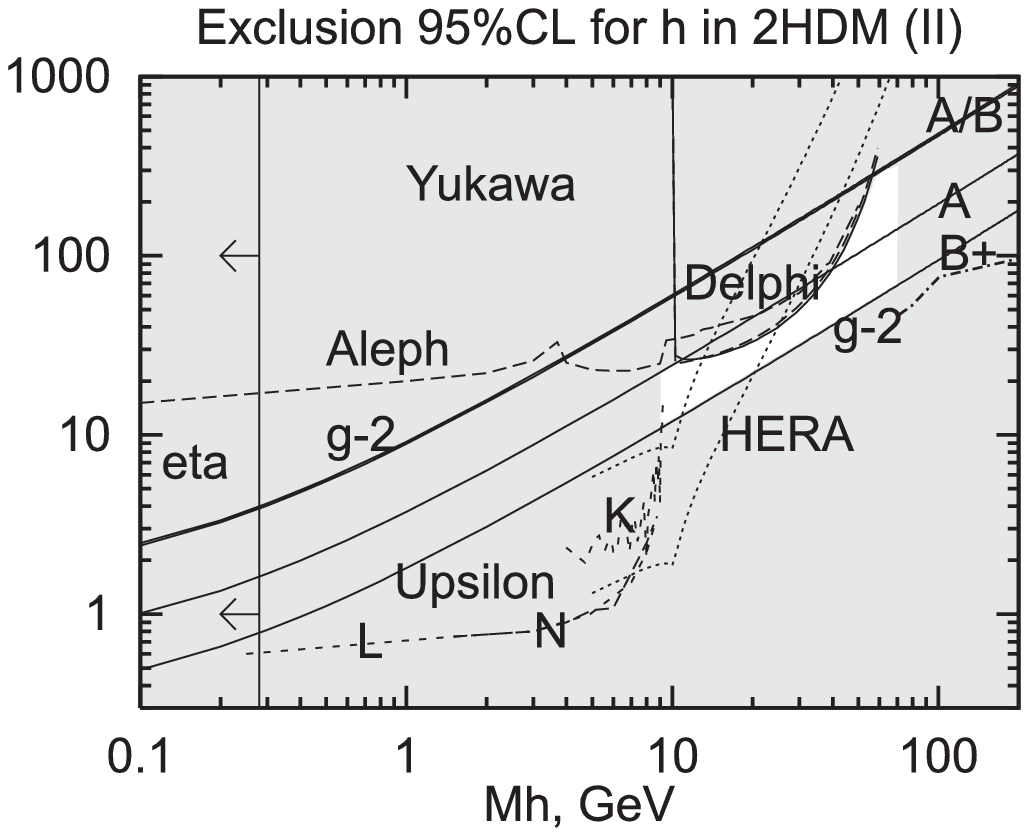}\\
\includegraphics{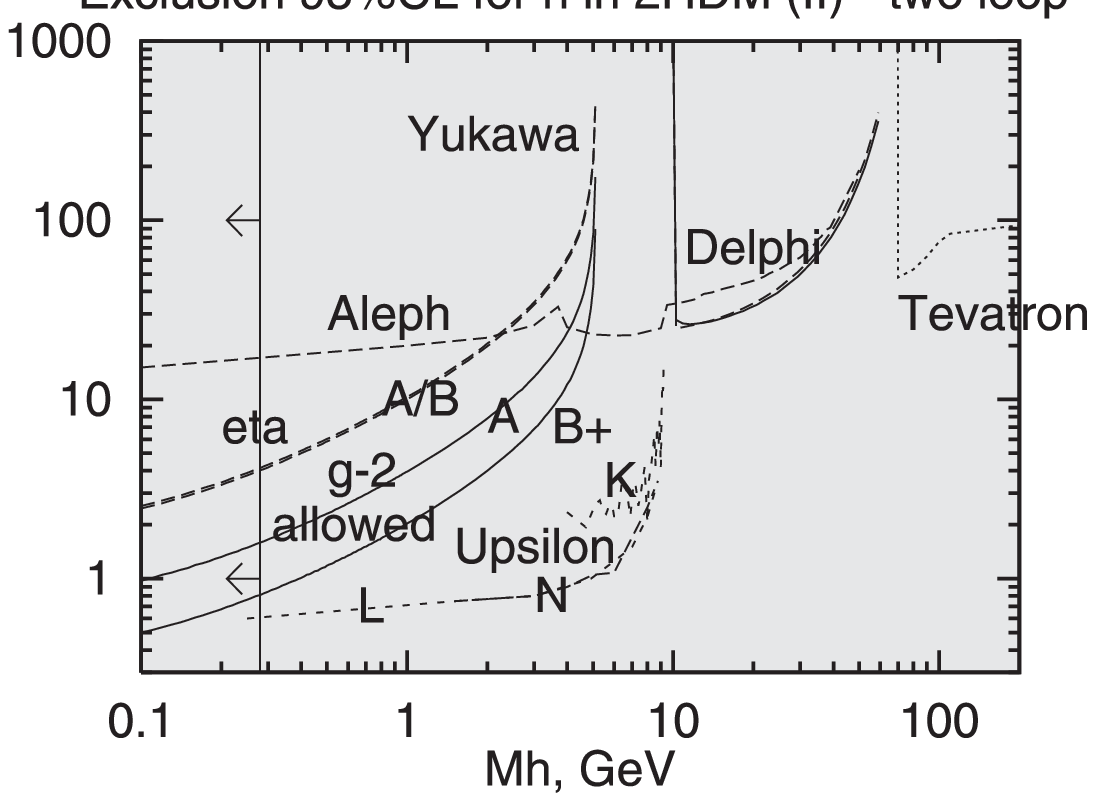}
\caption{\label{fig:6} 
Upper panel: one-loop result for $h$.   
Limits for the Yukawa coupling  of the scalar $h$  normalized to the 
SM value, $\chi_d^h$ (equal to $ \tan \beta $ for $\beta=\alpha $) 
as a function of the mass $M_h$. The constraint from the $ \eta $ decay 
 excludes 95\% CL  the mass below 0.28 GeV (vertical line),
 upper 95\% CL  limits from the Yukawa 
process (ALEPH measurements for pseudoscalar -  
dashed line, DELPHI data for scalar 
(pseudoscalar) -  solid (dashed) line). Upper  90\% CL limits from the 
$ \Upsilon \ra h/A \gamma $,  (K, N and L results, see text).
Results from  (K) are rescaled by a factor 2. 
The upper 95\% limits from the TEVATRON collider for $h$ and $A$ 
(rescaled by $\sqrt 2$ for masses below 130 GeV) are also shown 
(dot-dashed line). Allowed   
 $lim(95\%) $ bands for $h$ from the newest $g-2$ data 
for the muon: regions between the line   corresponding to  both  
A and B cases (denoted  A/B) and 
 lines giving  lower bounds  calculated for case A and case B ($lim_+(95\%)$)
 (denoted B$_+$),
 respectively. For a comparison  
the expected upper 95 \% CL limits from process $gg\ra h \ra \tau \tau $ at 
the $ep$ collider HERA are shown (dotted lines for lower and higher 
luminosity). Lower panel: sum of one- and two-loop results for $h$ 
(for $\beta=\alpha$).
Curves as in the upper panel.
}
\end{figure*}

\begin{figure*}[t]
\includegraphics{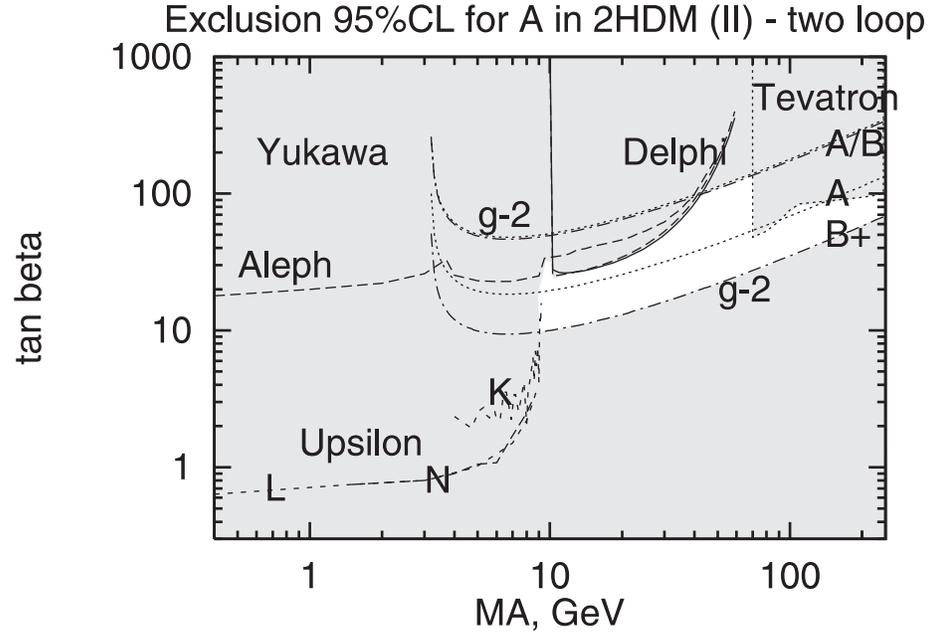}
\caption{\label{fig:7} 
Two-loop result for $A$. 
Limits for the Yukawa coupling  of the pseudoscalar $A$  normalized to the 
SM value, $\chi_d^A$ equal to $ \tan \beta $,   
as a function of the mass $M_A$. The allowed bands from $g-2$ data 
for A and B($lim_+(95\%)$) cases are shown
together with constraints from other experiments,
 details as in Figs.6. The white area is allowed by the all 
existing data.
}
\end{figure*}

\end{document}